\documentstyle[epsf,11pt]{article}
\newcommand{\PL}{{Phys.\ Lett.\/}}

\newcommand{\pr}{{Phys.\ Rev.\/}}

\newcommand{\np}{{Nucl.\ Phys.\/}}

\newcommand{\beq}{\begin{equation}}
\newcommand{\eeq}{\end{equation}}
\newcommand{\backn}{\begin{ack}}
\newcommand{\eackn}{\end{ack}}
\newcommand{\AmS}{{\protect\the\textfont2
  A\kern-.1667em\lower.5ex\hbox{M}\kern-.125emS}}

\title{Monopole Currents in $U(1)$ Lattice Gauge Theory: \\ A Comparison to an Effective Model \\ based on Dual Superconductivity \thanks{Supported by ``Jubil\"aumsfonds der Oesterreichischen Nationalbank'', Contract No.~4672}}

\author{Martin Zach, Manfried Faber, Wolfgang Kainz and Peter Skala\\
Institut f\"ur Kernphysik, Technische Universit\"at Wien, A--1040 Vienna, Austria}

\date{}

\begin{document}

\maketitle

\begin{abstract}
We compare $U(1)$ lattice gauge theory to an effective model of Maxwell and London equations. In the  effective model there is only one free parameter, the London penetration depth $\lambda$. It turns out that one can get good agreement between both models if one modifies the usual definition of magnetic monopole currents in $U(1)$ lattice gauge theory. This comparison also shows that already at small distances fluctuations of the occuring string are important. Further, we investigate the $\beta$-dependence of the penetration depth and determine the suppression of the monopole condensate in flux tubes.
\end{abstract}

\section{Introduction}

One of the most fundamental phenomena  which have to be explained by Quantum Chromodynamics (QCD) is the confinement of colour charges. Numerical simulations of QCD on a space-time lattice demonstrated the confinement phenomenon but could not yet clarify completely the mechanism  for the formation of the gluonic flux tube in the QCD vacuum. The most promising picture for this mechanism is the dual superconductor model \cite{mandelstam}: Dynamically generated colour magnetic monopole currents form a solenoid which squeezes the chromoelectric flux between quarks into a narrow flux tube. The energy of this flux tube increases linearly with its length and confinement is achieved.

Magnetic monopoles are topological excitations which condense in the confining phase. At strong coupling $U(1)$ lattice gauge theory has a confining phase and the monopole density is high, whereas it decreases exponentially above the phase transition to the Coulomb phase \cite{degrand}. Therefore it can be regarded as a prototype of a confining theory, and we will use it for studying some features of the confinement mechanism. $U(1)$ as a subgroup of $SU(N)$ is also important for investigating confinement in non-Abelian gauge theories, using the Abelian projection approach \cite{thooft}: After partially fixing a gauge one can define ``Abelian monopoles'' which could be responsible for confinement. The nature of these monopoles, however, remains obscure due to their dependence on the particular gauge that has been chosen. Interpreted in terms of \mbox{'t Hooft}-Polyakov-like monopoles they can possibly be regarded as regular extended objects which do have physical relevance \cite{vandersijs}. In any case, the identification of monopoles and their role in the microscopic mechanism of confinement is still an interesting open problem.

There have been many different approaches for describing the interaction of two charges in a confining medium. In the context of lattice gauge theory we would like to mention:
\begin{itemize}
\item calculation of flux tube profiles by Wilson-loop -- plaquette correlators,
\item application of London theory of dual superconductivity for the connection of field strength and monopole currents,
\item string picture of confinement (as exhibited in the Hamiltonian formulation of lattice gauge theories \cite{best}).
\end{itemize} 
The flux tube connecting two (colour) charges can be made visible by calculating observables like the energy or action density \cite{sommer}. For studying the confinement mechanism, however, it is also important to measure the electric field strength itself, rather than its square. This was done for $U(1)$ \cite{haymaker} and also for $SU(2)$ \cite{stefan}. The dual superconductor model assumes that the induced monopole currents obey a dual version of the London equation. Together with Ampere's law this determines the electric field distribution, which can be compared to the results of lattice simulations. In this way one obtains the ``London penetration length'' of the confinement vacuum. In ref.~\cite{haymaker} the persistent monopole current distribution in the presence of a charge pair is measured, and the results are interpreted as validity of fluxoid quantization known from flux tubes in ordinary superconductivity. A careful analysis of lattice results, however, shows that in the usual scheme a discrepancy appears between electric field and magnetic current distributions compared to the predictions based on an effective model of Maxwell and London equations. In this article we propose a modification of the identification scheme for monopoles. Within this scheme we get good agreement for electric flux and monopole current distributions between $U(1)$ lattice calculations and the effective model.

Lattice calculations also indicate that the width of the flux tube increases with the distance of sources. At least this is true in the case of finite temperature which is unavoidable when using smaller lattices. This means that in addition to the linear $\sigma r$ term in the potential there is also a Coulomb-like term. In the string picture of confinement this behaviour has been explained by string fluctuations \cite{luescher}. We investigate such string fluctuations within the dual superconductor picture. 

The effective model of coupled Maxwell-London equations has been solved in  ref. \cite{effmodel} in the continuum for a straight string. For performing accurate quantitative comparison between lattice simulation of $U(1)$ and an effective model based on the dual superconductor picture, it is necessary to take care of lattice effects. Therefore, we solve the effective model on the same lattice that we use for Monte Carlo calculations. In this way we gain control over anisotropy effects as well as finite size effects (using periodic boundary conditions). It will also be possible to consider the finite temperature in our $U(1)$ simulations which is due to the finite time extension of the lattice.

In section 2 of this article we discuss our modification of identifying magnetic monopoles and determine electric field and magnetic current distributions around a pair of charges. Finally, we show the behaviour of the monopole condensate in external electric fields. In section 3 we perform a comparison between $U(1)$ lattice gauge theory and the results of the effective model and investigate the dependence of the penetration depth $\lambda$ on the coupling.

\section{Electric field strength and mono\-pole currents in $U(1)$}

We examine the electric field and monopole current distribution in the strong coupling phase in the presence of a pair of electric charges. A Euclidean $8^3 \times 4$ lattice with periodic boundary conditions is suitable to compare lattice calculations and effective model with sufficient accuracy and to reveal also tiny effects. $U(1)$ gauge degrees of freedom are represented by link variables $U_{\mu}(x)=e^{i\theta_{\mu}(x)}$, where the phases $\theta_{\mu}(x)\in(-\pi,\pi]$ correspond to $e a A_{\mu}(x)$ in the continuum limit ($e$ is the coupling constant and $a$ the lattice spacing). For the simulation we use the standard Wilson action
\begin{equation} \label{action}
S = \beta \sum_{x, \mu < \nu} [1-{\rm cos}\theta_{\mu\nu}(x)],
\end{equation}
where $\beta=\frac{1}{e^2}$ is the inverse coupling and $\theta_{\mu\nu}(x)\in(-\pi,\pi]$ is the phase of the plaquette variable
\begin{equation} \label{plaqu}
U_{\mu\nu}(x)=e^{i\theta_{\mu\nu}(x)}=U_{\mu}(x)U_{\nu}(x+\hat{\mu})U^\dagger_{\mu}(x+\hat{\nu})U^\dagger_{\nu}(x).
\end{equation}
Static charges can be represented by Polyakov loops $L(\vec{r})=\prod_{k=1}^{N_{t}}U_{\mu=0}[x=(\vec r,ka)]$. We are able to determine expectation values of physical observables ${\cal O}$ around a charge pair at a distance $d$ by the correlation function 
\begin{equation} \label{correlation}
\langle {\cal O}(x) \rangle_{Q\bar{Q}} = \frac{\langle L(0) L^\ast(d) {\cal O}(x) \rangle}{\langle L(0) L^\ast(d) \rangle} - \langle {\cal O} \rangle,
\end{equation}
where the angle brackets denote the evaluation of the path integral. Let us recall now the definition of electric fields and explain our modification in identifying magnetic currents. These definitions turn out to be essential for the validity of a lattice version of the Gauss law and fluxoid quantization.

In the continuum the phase $\theta_{\mu\nu}$ of a Wilson loop around an infinitely small square of size $a^2$ in the $\mu \nu$-plane defines the field strength $F_{\mu\nu}$ by $a^2 e F_{\mu\nu} = \theta_{\mu\nu}$. On a finite plaquette, however, two (nearly) equivalent plaquette angles with values $\pm \pi (\mp \varepsilon)$ would correspond to two drastically different values of the field strength $F_{\mu\nu} = \pm \pi / (a^2 e)$. There is a large class of functions which fulfil the requirements of the correct continuum limit and $2\pi$-periodicity. Therefore we demand that the Gauss law connects the field strength with external charges $\pm e$ which we put in the vacuum.

A derivation of the Gauss law shows that the appropriate field strength definition depends on the action used. Since the path integral is invariant under translations of link phases $\theta_{\mu}(x)$, we perform a variation of the angle \mbox{$\theta_4(\vec{r},x_4)$} in the expectation value of a pair of Polyakov loops:
\begin{equation}\label{variation}
\frac{\delta}{\delta \theta_4(x)} \int D[U] e^{-S[U]} L(\vec{r}_+)
L^{\ast}(\vec{r}_-) = 0.
\end{equation}
For the Wilson action this results in
\begin{equation} \label{Gauss_law}
\sum_{i=1 \atop (\theta_4(x) \in \Box_i)}^6 (\pm) \; \frac{\langle L(\vec{r}_+)  L^\ast(\vec{r}_-) \frac{1}{e} \sin \theta_{\Box_i} \rangle} {\langle L(\vec{r}_+)  L^\ast(\vec{r}_-) \rangle} =
i e \left[ \delta_{\vec{r},\vec{r}_+} - \delta_{\vec{r},\vec{r}_-} \right],
\end{equation}
where the sum runs over the six space-time plaquettes including the regarded link $\theta_4(x)$, and the sign accounts for the direction of the link in the plaquette $\Box_i$. If we identify
\begin{equation} \label{divergence}
a^3 {\rm div} {\vec{E}}(x) = \sum_{i=1 \atop (\theta_4(x) \in \Box_i)}^6 (\pm) \; \frac{1}{e} \sin \theta_{\Box_i},
\end{equation}
we get for the Wilson action the natural Euclidean field strength definition
\begin{equation} \label{field_strength}
a^2 e F_{\mu\nu} = \sin \theta_{\mu\nu}.
\end{equation}
From the validity of equ.~(\ref{Gauss_law}) it can be seen that for the Euclidean metric the expectation values of the charge density \cite{charge} as well as electric fields turn out to be imaginary. The corresponding observables in Minkowski space are real as they should.

DeGrand and Toussaint \cite{degrand} suggested to identify magnetic monopoles as ends of Dirac strings \cite{dirac}. If the sum over link phases around an elementary plaquette exceeds $\pm \pi$ one defines that a Dirac string is crossing this plaquette. In the case that the net number of Dirac strings through the surface of a three-dimensional cube is nonzero a monopole or antimonopole is located in the center of this cube. Whereas the location of a Dirac string is gauge dependent and therefore cannot have a physical meaning, the monopole number inside a cube is a gauge invariant quantity. This is easily seen by expressing the integer valued monopole current $n_\mu$ through the sum of six gauge invariant plaquette phases which form the boundary of a three-dimensional cube:
\begin{equation}\label{monodt}
\sum_{\Box_i \in {\rm cube}(\nu \rho \sigma)} \theta_{\Box_i} = 2 \pi n^{\mu} \varepsilon_{\mu \nu \rho \sigma},
\end{equation}
which is equivalent to counting Dirac strings. Each Dirac string carries a quantized magnetic flux of $2 \pi / e$, therefore the magnetic current in terms of the electric coupling reads
\begin{equation}\label{dt}
J_{m\,\rm DT}^{\mu} = \frac{2 \pi}{e} n^{\mu}.
\end{equation}
We suggest to modify this definition and to use instead of (\ref{monodt}) and (\ref{dt}) the dual Maxwell equations for the identification of monopole currents. These read on the lattice
\begin{equation} \label{monop}
a^2 \sum_{\Box_i \in {\rm cube}(\nu \rho \sigma)} F_{\Box_i} = J_m^\mu \varepsilon_{\mu \nu \rho \sigma},
\end{equation}
with $F_{\Box_i}$ according to our field strength definition (\ref{field_strength}). Monopole currents defined in this way form closed loops on the dual lattice and obey therefore a current conservation law. However, it is no longer true that the monopole charge in each cube is an integer multiple of $2 \pi / e$. Only if a separated monopole is investigated in the classical limit ($S \to min$) the flux through a large surface enclosing it turns out to be quantized due to $\sin \theta \to \theta$.

With the definitions (\ref{field_strength}) and (\ref{monop}) we determine the distribution of the electric field strength $E$ and the magnetic currents $J_m$ in the confinement phase ($\beta=0.96$) for a pair of static charges $\pm e$ at distances from $1a$ to $4a$. For our ``measurements'' we take $20\,000$ gauge field configurations which have been generated by a standard Metropolis algorithm and are separated from each other by 200 complete updates. The errorbars shown in the figures indicate the jack knife error, which is around 50 \% higher than the naive statistical one. As expected, the monopole currents behave like a coil, squeezing the electric field into a narrow flux tube.  The magnetic monopoles form left-handed currents around the electric flux lines.

\begin{figure}
\centerline{% GNUPLOT: LaTeX picture with Postscript
\setlength{\unitlength}{0.1bp}
\special{!
%!PS-Adobe-2.0
%%Creator: gnuplot
%%DocumentFonts: Helvetica
%%BoundingBox: 50 50 518 403
%%Pages: (atend)
%%EndComments
/gnudict 40 dict def
gnudict begin
/Color false def
/Solid false def
/gnulinewidth 2.000 def
/vshift -33 def
/dl {10 mul} def
/hpt 31.5 def
/vpt 31.5 def
/M {moveto} bind def
/L {lineto} bind def
/R {rmoveto} bind def
/V {rlineto} bind def
/vpt2 vpt 2 mul def
/hpt2 hpt 2 mul def
/Lshow { currentpoint stroke M
  0 vshift R show } def
/Rshow { currentpoint stroke M
  dup stringwidth pop neg vshift R show } def
/Cshow { currentpoint stroke M
  dup stringwidth pop -2 div vshift R show } def
/DL { Color {setrgbcolor Solid {pop []} if 0 setdash }
 {pop pop pop Solid {pop []} if 0 setdash} ifelse } def
/BL { stroke gnulinewidth 2 mul setlinewidth } def
/AL { stroke gnulinewidth 2 div setlinewidth } def
/PL { stroke gnulinewidth setlinewidth } def
/LTb { BL [] 0 0 0 DL } def
/LTa { AL [1 dl 2 dl] 0 setdash 0 0 0 setrgbcolor } def
/LT0 { PL [] 0 1 0 DL } def
/LT1 { PL [4 dl 2 dl] 0 0 1 DL } def
/LT2 { PL [2 dl 3 dl] 1 0 0 DL } def
/LT3 { PL [1 dl 1.5 dl] 1 0 1 DL } def
/LT4 { PL [5 dl 2 dl 1 dl 2 dl] 0 1 1 DL } def
/LT5 { PL [4 dl 3 dl 1 dl 3 dl] 1 1 0 DL } def
/LT6 { PL [2 dl 2 dl 2 dl 4 dl] 0 0 0 DL } def
/LT7 { PL [2 dl 2 dl 2 dl 2 dl 2 dl 4 dl] 1 0.3 0 DL } def
/LT8 { PL [2 dl 2 dl 2 dl 2 dl 2 dl 2 dl 2 dl 4 dl] 0.5 0.5 0.5 DL } def
/P { stroke [] 0 setdash
  currentlinewidth 2 div sub M
  0 currentlinewidth V stroke } def
/D { stroke [] 0 setdash 2 copy vpt add M
  hpt neg vpt neg V hpt vpt neg V
  hpt vpt V hpt neg vpt V closepath stroke
  P } def
/A { stroke [] 0 setdash vpt sub M 0 vpt2 V
  currentpoint stroke M
  hpt neg vpt neg R hpt2 0 V stroke
  } def
/B { stroke [] 0 setdash 2 copy exch hpt sub exch vpt add M
  0 vpt2 neg V hpt2 0 V 0 vpt2 V
  hpt2 neg 0 V closepath stroke
  P } def
/C { stroke [] 0 setdash exch hpt sub exch vpt add M
  hpt2 vpt2 neg V currentpoint stroke M
  hpt2 neg 0 R hpt2 vpt2 V stroke } def
/T { stroke [] 0 setdash 2 copy vpt 1.12 mul add M
  hpt neg vpt -1.62 mul V
  hpt 2 mul 0 V
  hpt neg vpt 1.62 mul V closepath stroke
  P  } def
/S { 2 copy A C} def
end
}
\begin{picture}(2339,1511)(0,0)
\special{"
gnudict begin
gsave
50 50 translate
0.100 0.100 scale
0 setgray
/Helvetica findfont 100 scalefont setfont
newpath
-500.000000 -500.000000 translate
LTa
600 1259 M
1556 0 V
1378 251 M
0 1209 V
LTb
600 251 M
63 0 V
1493 0 R
-63 0 V
600 452 M
63 0 V
1493 0 R
-63 0 V
600 654 M
63 0 V
1493 0 R
-63 0 V
600 855 M
63 0 V
1493 0 R
-63 0 V
600 1057 M
63 0 V
1493 0 R
-63 0 V
600 1259 M
63 0 V
1493 0 R
-63 0 V
600 1460 M
63 0 V
1493 0 R
-63 0 V
711 251 M
0 63 V
0 1146 R
0 -63 V
933 251 M
0 63 V
0 1146 R
0 -63 V
1156 251 M
0 63 V
0 1146 R
0 -63 V
1378 251 M
0 63 V
0 1146 R
0 -63 V
1600 251 M
0 63 V
0 1146 R
0 -63 V
1823 251 M
0 63 V
0 1146 R
0 -63 V
2045 251 M
0 63 V
0 1146 R
0 -63 V
600 251 M
1556 0 V
0 1209 V
-1556 0 V
600 251 L
LT0
711 1252 M
0 33 V
-31 -33 R
62 0 V
-62 33 R
62 0 V
191 105 R
0 33 V
-31 -33 R
62 0 V
-62 33 R
62 0 V
1156 325 M
0 33 V
-31 -33 R
62 0 V
-62 33 R
62 0 V
191 175 R
0 47 V
-31 -47 R
62 0 V
-62 47 R
62 0 V
1600 325 M
0 33 V
-31 -33 R
62 0 V
-62 33 R
62 0 V
192 1032 R
0 33 V
-31 -33 R
62 0 V
-62 33 R
62 0 V
191 -171 R
0 33 V
-31 -33 R
62 0 V
-62 33 R
62 0 V
600 1259 M
0 10 V
222 0 V
0 -10 V
-222 0 V
222 0 R
0 148 V
223 0 V
0 -148 V
-223 0 V
223 0 R
0 -918 V
222 0 V
0 918 V
-222 0 V
222 0 R
0 -703 V
222 0 V
0 703 V
-222 0 V
222 0 R
0 -918 V
222 0 V
0 918 V
-222 0 V
222 0 R
0 148 V
223 0 V
0 -148 V
-223 0 V
223 0 R
0 10 V
222 0 V
0 -10 V
-222 0 V
stroke
grestore
end
showpage
}
\put(1378,51){\makebox(0,0){zeta [a]}}
\put(200,855){%
\special{ps: gsave currentpoint currentpoint translate
270 rotate neg exch neg exch translate}%
\makebox(0,0)[b]{\shortstack{$E \;[\rm a^{-2}]$}}%
\special{ps: currentpoint grestore moveto}%
}
\put(2045,151){\makebox(0,0){3}}
\put(1823,151){\makebox(0,0){2}}
\put(1600,151){\makebox(0,0){1}}
\put(1378,151){\makebox(0,0){0}}
\put(1156,151){\makebox(0,0){-1}}
\put(933,151){\makebox(0,0){-2}}
\put(711,151){\makebox(0,0){-3}}
\put(540,1460){\makebox(0,0)[r]{0.1}}
\put(540,1259){\makebox(0,0)[r]{0}}
\put(540,1057){\makebox(0,0)[r]{-0.1}}
\put(540,855){\makebox(0,0)[r]{-0.2}}
\put(540,654){\makebox(0,0)[r]{-0.3}}
\put(540,452){\makebox(0,0)[r]{-0.4}}
\put(540,251){\makebox(0,0)[r]{-0.5}}
\end{picture}% GNUPLOT: LaTeX picture with Postscript
\setlength{\unitlength}{0.1bp}
\special{!
%!PS-Adobe-2.0
%%Creator: gnuplot
%%DocumentFonts: Helvetica
%%BoundingBox: 50 50 518 403
%%Pages: (atend)
%%EndComments
/gnudict 40 dict def
gnudict begin
/Color false def
/Solid false def
/gnulinewidth 2.000 def
/vshift -33 def
/dl {10 mul} def
/hpt 31.5 def
/vpt 31.5 def
/M {moveto} bind def
/L {lineto} bind def
/R {rmoveto} bind def
/V {rlineto} bind def
/vpt2 vpt 2 mul def
/hpt2 hpt 2 mul def
/Lshow { currentpoint stroke M
  0 vshift R show } def
/Rshow { currentpoint stroke M
  dup stringwidth pop neg vshift R show } def
/Cshow { currentpoint stroke M
  dup stringwidth pop -2 div vshift R show } def
/DL { Color {setrgbcolor Solid {pop []} if 0 setdash }
 {pop pop pop Solid {pop []} if 0 setdash} ifelse } def
/BL { stroke gnulinewidth 2 mul setlinewidth } def
/AL { stroke gnulinewidth 2 div setlinewidth } def
/PL { stroke gnulinewidth setlinewidth } def
/LTb { BL [] 0 0 0 DL } def
/LTa { AL [1 dl 2 dl] 0 setdash 0 0 0 setrgbcolor } def
/LT0 { PL [] 0 1 0 DL } def
/LT1 { PL [4 dl 2 dl] 0 0 1 DL } def
/LT2 { PL [2 dl 3 dl] 1 0 0 DL } def
/LT3 { PL [1 dl 1.5 dl] 1 0 1 DL } def
/LT4 { PL [5 dl 2 dl 1 dl 2 dl] 0 1 1 DL } def
/LT5 { PL [4 dl 3 dl 1 dl 3 dl] 1 1 0 DL } def
/LT6 { PL [2 dl 2 dl 2 dl 4 dl] 0 0 0 DL } def
/LT7 { PL [2 dl 2 dl 2 dl 2 dl 2 dl 4 dl] 1 0.3 0 DL } def
/LT8 { PL [2 dl 2 dl 2 dl 2 dl 2 dl 2 dl 2 dl 4 dl] 0.5 0.5 0.5 DL } def
/P { stroke [] 0 setdash
  currentlinewidth 2 div sub M
  0 currentlinewidth V stroke } def
/D { stroke [] 0 setdash 2 copy vpt add M
  hpt neg vpt neg V hpt vpt neg V
  hpt vpt V hpt neg vpt V closepath stroke
  P } def
/A { stroke [] 0 setdash vpt sub M 0 vpt2 V
  currentpoint stroke M
  hpt neg vpt neg R hpt2 0 V stroke
  } def
/B { stroke [] 0 setdash 2 copy exch hpt sub exch vpt add M
  0 vpt2 neg V hpt2 0 V 0 vpt2 V
  hpt2 neg 0 V closepath stroke
  P } def
/C { stroke [] 0 setdash exch hpt sub exch vpt add M
  hpt2 vpt2 neg V currentpoint stroke M
  hpt2 neg 0 R hpt2 vpt2 V stroke } def
/T { stroke [] 0 setdash 2 copy vpt 1.12 mul add M
  hpt neg vpt -1.62 mul V
  hpt 2 mul 0 V
  hpt neg vpt 1.62 mul V closepath stroke
  P  } def
/S { 2 copy A C} def
end
}
\begin{picture}(2139,1511)(200,0)
\special{"
gnudict begin
gsave
50 50 translate
0.100 0.100 scale
0 setgray
/Helvetica findfont 100 scalefont setfont
newpath
-500.000000 -500.000000 translate
LTa
600 251 M
1556 0 V
-778 0 R
0 1209 V
LTb
600 251 M
63 0 V
1493 0 R
-63 0 V
600 453 M
63 0 V
1493 0 R
-63 0 V
600 654 M
63 0 V
1493 0 R
-63 0 V
600 856 M
63 0 V
1493 0 R
-63 0 V
600 1057 M
63 0 V
1493 0 R
-63 0 V
600 1259 M
63 0 V
1493 0 R
-63 0 V
600 1460 M
63 0 V
1493 0 R
-63 0 V
711 251 M
0 63 V
0 1146 R
0 -63 V
933 251 M
0 63 V
0 1146 R
0 -63 V
1156 251 M
0 63 V
0 1146 R
0 -63 V
1378 251 M
0 63 V
0 1146 R
0 -63 V
1600 251 M
0 63 V
0 1146 R
0 -63 V
1823 251 M
0 63 V
0 1146 R
0 -63 V
2045 251 M
0 63 V
0 1146 R
0 -63 V
600 251 M
1556 0 V
0 1209 V
-1556 0 V
600 251 L
LT0
711 291 M
0 75 V
680 291 M
62 0 V
-62 75 R
62 0 V
191 64 R
0 76 V
902 430 M
62 0 V
-62 76 R
62 0 V
192 818 R
0 76 V
-31 -76 R
62 0 V
-62 76 R
62 0 V
191 -392 R
0 107 V
-31 -107 R
62 0 V
-62 107 R
62 0 V
191 209 R
0 76 V
-31 -76 R
62 0 V
-62 76 R
62 0 V
1823 430 M
0 76 V
-31 -76 R
62 0 V
-62 76 R
62 0 V
2045 291 M
0 75 V
-31 -75 R
62 0 V
-62 75 R
62 0 V
600 251 M
0 78 V
222 0 V
0 -78 V
-222 0 V
222 0 R
0 217 V
223 0 V
0 -217 V
-223 0 V
223 0 R
0 1111 V
222 0 V
0 -1111 V
-222 0 V
222 0 R
0 811 V
222 0 V
0 -811 V
-222 0 V
222 0 R
0 1111 V
222 0 V
0 -1111 V
-222 0 V
222 0 R
0 217 V
223 0 V
0 -217 V
-223 0 V
223 0 R
0 78 V
222 0 V
0 -78 V
-222 0 V
stroke
grestore
end
showpage
}
\put(1378,51){\makebox(0,0){zeta [a]}}
\put(200,855){%
\special{ps: gsave currentpoint currentpoint translate
270 rotate neg exch neg exch translate}%
\makebox(0,0)[b]{\shortstack{rot $J_m \;[\rm a^{-4}]$}}%
\special{ps: currentpoint grestore moveto}%
}
\put(2045,151){\makebox(0,0){3}}
\put(1823,151){\makebox(0,0){2}}
\put(1600,151){\makebox(0,0){1}}
\put(1378,151){\makebox(0,0){0}}
\put(1156,151){\makebox(0,0){-1}}
\put(933,151){\makebox(0,0){-2}}
\put(711,151){\makebox(0,0){-3}}
\put(540,1460){\makebox(0,0)[r]{1.2}}
\put(540,1259){\makebox(0,0)[r]{1}}
\put(540,1057){\makebox(0,0)[r]{0.8}}
\put(540,856){\makebox(0,0)[r]{0.6}}
\put(540,654){\makebox(0,0)[r]{0.4}}
\put(540,453){\makebox(0,0)[r]{0.2}}
\put(540,251){\makebox(0,0)[r]{0}}
\end{picture}}
\caption{\label{axis}Longitudinal profile of the electric field and the curl of the magnetic current on the $Q\bar{Q}$-axis for a distance $d=3a$ of charges: Both quantities give a decreasing signal in the middle between charges ($\rm zeta=0$).} 
\end{figure}
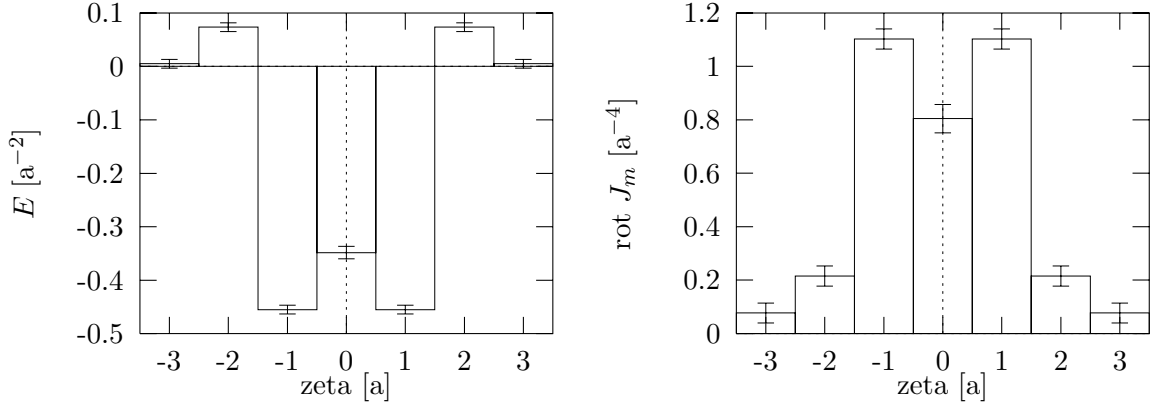

\begin{figure}
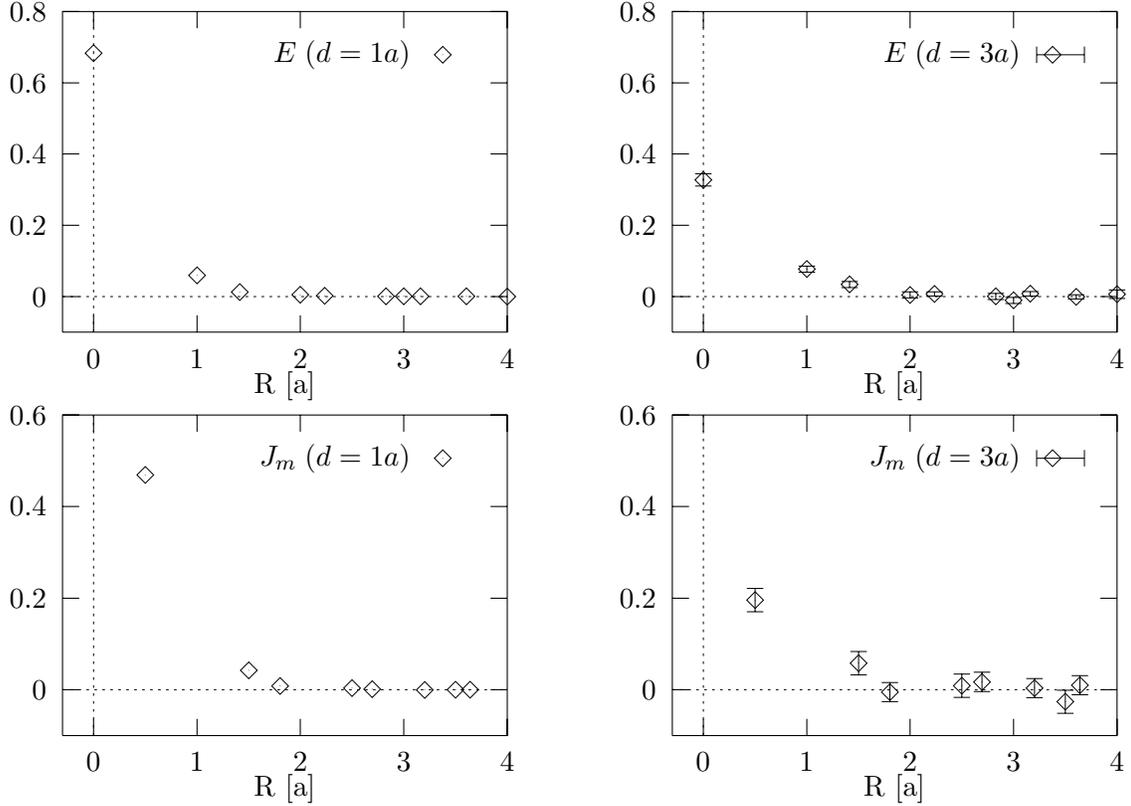

\centerline{\input{e1.tex}\input{e3.tex}\hspace{1cm}}
\centerline{\input{jm1.tex}\input{jm3.tex}\hspace{1cm}}
\caption{\label{transv}Transverse profile of electric field strength and monopole currents for distances $d=1a$ and $d=3a$, in the symmetry plane between charges (zeta $=0$) : For $d=1a$ the flux tube is concentrated on the link connecting the charges, for $d=3a$ it spreads out significantly which can also be seen by the decreasing monopole currents.}
\end{figure}

In Fig.~\ref{axis} we show the electric field and the curl of the magnetic current on the $Q\bar{Q}$-axis for a distance $d=3a$ of charges. The absolute values of both quantities decrease clearly in the middle between charges. This is an indication for string fluctuations as will be discussed in the next section. A transverse profile of electric field strength and monopole currents, which characterizes the thickness of the flux tube, is shown in Fig.~\ref{transv} for distances $d=1a$ and $d=3a$ of charges.

\begin{figure}
\centerline{% GNUPLOT: LaTeX picture with Postscript
\setlength{\unitlength}{0.1bp}
\special{!
%!PS-Adobe-2.0
%%Creator: gnuplot
%%DocumentFonts: Helvetica
%%BoundingBox: 50 50 518 403
%%Pages: (atend)
%%EndComments
/gnudict 40 dict def
gnudict begin
/Color false def
/Solid false def
/gnulinewidth 2.000 def
/vshift -33 def
/dl {10 mul} def
/hpt 31.5 def
/vpt 31.5 def
/M {moveto} bind def
/L {lineto} bind def
/R {rmoveto} bind def
/V {rlineto} bind def
/vpt2 vpt 2 mul def
/hpt2 hpt 2 mul def
/Lshow { currentpoint stroke M
  0 vshift R show } def
/Rshow { currentpoint stroke M
  dup stringwidth pop neg vshift R show } def
/Cshow { currentpoint stroke M
  dup stringwidth pop -2 div vshift R show } def
/DL { Color {setrgbcolor Solid {pop []} if 0 setdash }
 {pop pop pop Solid {pop []} if 0 setdash} ifelse } def
/BL { stroke gnulinewidth 2 mul setlinewidth } def
/AL { stroke gnulinewidth 2 div setlinewidth } def
/PL { stroke gnulinewidth setlinewidth } def
/LTb { BL [] 0 0 0 DL } def
/LTa { AL [1 dl 2 dl] 0 setdash 0 0 0 setrgbcolor } def
/LT0 { PL [] 0 1 0 DL } def
/LT1 { PL [4 dl 2 dl] 0 0 1 DL } def
/LT2 { PL [2 dl 3 dl] 1 0 0 DL } def
/LT3 { PL [1 dl 1.5 dl] 1 0 1 DL } def
/LT4 { PL [5 dl 2 dl 1 dl 2 dl] 0 1 1 DL } def
/LT5 { PL [4 dl 3 dl 1 dl 3 dl] 1 1 0 DL } def
/LT6 { PL [2 dl 2 dl 2 dl 4 dl] 0 0 0 DL } def
/LT7 { PL [2 dl 2 dl 2 dl 2 dl 2 dl 4 dl] 1 0.3 0 DL } def
/LT8 { PL [2 dl 2 dl 2 dl 2 dl 2 dl 2 dl 2 dl 4 dl] 0.5 0.5 0.5 DL } def
/P { stroke [] 0 setdash
  currentlinewidth 2 div sub M
  0 currentlinewidth V stroke } def
/D { stroke [] 0 setdash 2 copy vpt add M
  hpt neg vpt neg V hpt vpt neg V
  hpt vpt V hpt neg vpt V closepath stroke
  P } def
/A { stroke [] 0 setdash vpt sub M 0 vpt2 V
  currentpoint stroke M
  hpt neg vpt neg R hpt2 0 V stroke
  } def
/B { stroke [] 0 setdash 2 copy exch hpt sub exch vpt add M
  0 vpt2 neg V hpt2 0 V 0 vpt2 V
  hpt2 neg 0 V closepath stroke
  P } def
/C { stroke [] 0 setdash exch hpt sub exch vpt add M
  hpt2 vpt2 neg V currentpoint stroke M
  hpt2 neg 0 R hpt2 vpt2 V stroke } def
/T { stroke [] 0 setdash 2 copy vpt 1.12 mul add M
  hpt neg vpt -1.62 mul V
  hpt 2 mul 0 V
  hpt neg vpt 1.62 mul V closepath stroke
  P  } def
/S { 2 copy A C} def
end
}
\begin{picture}(2339,1511)(0,0)
\special{"
gnudict begin
gsave
50 50 translate
0.100 0.100 scale
0 setgray
/Helvetica findfont 100 scalefont setfont
newpath
-500.000000 -500.000000 translate
LTa
480 553 M
1676 0 V
480 251 M
0 1209 V
LTb
480 251 M
63 0 V
1613 0 R
-63 0 V
480 553 M
63 0 V
1613 0 R
-63 0 V
480 856 M
63 0 V
1613 0 R
-63 0 V
480 1158 M
63 0 V
1613 0 R
-63 0 V
480 1460 M
63 0 V
1613 0 R
-63 0 V
480 251 M
0 63 V
0 1146 R
0 -63 V
899 251 M
0 63 V
0 1146 R
0 -63 V
1318 251 M
0 63 V
0 1146 R
0 -63 V
1737 251 M
0 63 V
0 1146 R
0 -63 V
2156 251 M
0 63 V
0 1146 R
0 -63 V
480 251 M
1676 0 V
0 1209 V
-1676 0 V
480 251 L
LT0
1913 1297 C
690 1174 C
1109 636 C
1235 599 C
1528 567 C
1608 563 C
1822 562 C
1947 552 C
2005 560 C
1853 1297 M
180 0 V
-180 31 R
0 -62 V
180 62 R
0 -62 V
690 1164 M
0 20 V
-31 -20 R
62 0 V
-62 20 R
62 0 V
1109 626 M
0 20 V
-31 -20 R
62 0 V
-62 20 R
62 0 V
95 -54 R
0 14 V
-31 -14 R
62 0 V
-62 14 R
62 0 V
262 -49 R
0 20 V
-31 -20 R
62 0 V
-62 20 R
62 0 V
49 -21 R
0 14 V
-31 -14 R
62 0 V
-62 14 R
62 0 V
183 -15 R
0 14 V
-31 -14 R
62 0 V
-62 14 R
62 0 V
94 -27 R
0 20 V
-31 -20 R
62 0 V
-62 20 R
62 0 V
27 -9 R
0 14 V
-31 -14 R
62 0 V
-62 14 R
62 0 V
LT0
1913 1197 D
776 327 D
1142 497 D
1369 534 D
1548 549 D
1701 549 D
1962 556 D
1962 555 D
2076 553 D
1853 1197 M
180 0 V
-180 31 R
0 -62 V
180 62 R
0 -62 V
776 321 M
0 12 V
745 321 M
62 0 V
-62 12 R
62 0 V
335 160 R
0 8 V
-31 -8 R
62 0 V
-62 8 R
62 0 V
196 27 R
0 12 V
-31 -12 R
62 0 V
-62 12 R
62 0 V
148 5 R
0 8 V
-31 -8 R
62 0 V
-62 8 R
62 0 V
122 -8 R
0 8 V
-31 -8 R
62 0 V
-62 8 R
62 0 V
230 -3 R
0 12 V
-31 -12 R
62 0 V
-62 12 R
62 0 V
-31 -12 R
0 9 V
-31 -9 R
62 0 V
-62 9 R
62 0 V
83 -10 R
0 9 V
-31 -9 R
62 0 V
-62 9 R
62 0 V
stroke
grestore
end
showpage
}
\put(1793,1177){\makebox(0,0)[r]{$<J_{m,t}^2>$}}
\put(1793,1317){\makebox(0,0)[r]{$<J_{m,i}^2>$}}
\put(1318,51){\makebox(0,0){R [a]}}
\put(2156,151){\makebox(0,0){4}}
\put(1737,151){\makebox(0,0){3}}
\put(1318,151){\makebox(0,0){2}}
\put(899,151){\makebox(0,0){1}}
\put(480,151){\makebox(0,0){0}}
\put(420,1460){\makebox(0,0)[r]{1.5}}
\put(420,1158){\makebox(0,0)[r]{1}}
\put(420,856){\makebox(0,0)[r]{0.5}}
\put(420,553){\makebox(0,0)[r]{0}}
\put(420,251){\makebox(0,0)[r]{-0.5}}
\end{picture}\hspace{1cm}}
\caption{\label{condensate}The square of spatial ($<J_{m,i}^2>$) and temporal ($<J_{m,t}^2>$) monopole links in a transverse profile of the flux tube for a distance $d=2a$ of charges.}
\end{figure}
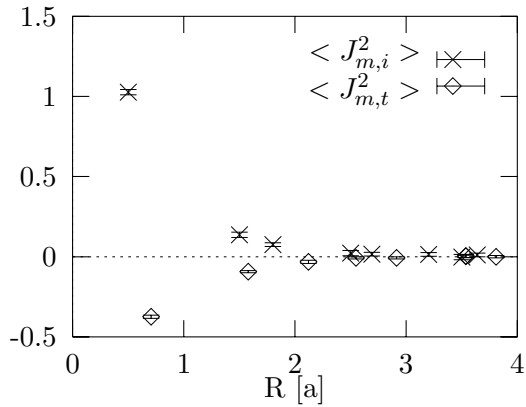

It is also interesting to look at the expectation values of the squared monopole currents in the region of the flux tube. The vacuum of the confinement phase without sources is crowded with randomly distributed current loops which are ``aligned'' by an external electric field. Calculations on the Euclidean lattice show that both spatial and temporal components of the squared monopole current are suppressed in the flux tube, i.e. they are negative compared to their vacuum expectation value. One has to take into account, however, that due to the definition (\ref{monop}) the spatial monopole currents are again purely imaginary, like the electric field strength, in the Euclidean metric. Therefore, the sign has to be changed when going back into Minkowski space. Consequently, the expectation values of the squared components of the monopole current behave differently in regions of electric fields, as shown in Fig.~\ref{condensate}: The temporal component ($=$ the monopole density) is suppressed compared to the vacuum, whereas the square of the spatial components is increased by the alignment process. The situation is the same as for the calculation of electric and magnetic energy density \cite{sommer}, where the sign has to be changed for the square of the electric field strength. The expectation value of $\vec{E}^2$ in the flux tube is positive, $\vec{B}^2$ turns out to be negative with respect to the vacuum. The observed decrease in the monopole density shows a further physical connection to a real superconductor, where the Cooper pair condensate is suppressed in regions of external magnetic fields. 
 
\section{A Classical Effective Model}

For a pair of static point charges $\rho(\vec{r}) = e [\delta(\vec{r} - \vec{r}_+) - \delta(\vec{r} - \vec{r}_-)]$ the coupled Maxwell-London equations in a dual superconductor read
{\begin{eqnarray}
\label{Gauss}
{\rm div}\vec{E}({\vec r}) &=& \rho(\vec{r}),\\
\label{Ampere}
{\rm rot}\vec{E}(\vec{r}) &=& -\vec{J_m}(\vec{r}),\\
\label{London}
\vec{E}(\vec{r}) &=& \lambda^2 {\rm rot}\vec{J_m}(\vec{r}).
\end{eqnarray}}

There is a discrepancy between the Gauss law and the London equation. Therefore, we have to introduce a dual Dirac string \cite{nambu} carrying a quantum of electric flux
\begin{equation} \label{string}
\vec{\cal E}(\vec{r}) = e \int_{l} d\vec{r}'\quad \delta^{(3)} (\vec{r}-\vec{r}'),
\end{equation}
where $l$ is an arbitrary line connecting the charges. Instead of the London equation (\ref{London}) we write
\begin{equation}
\label{fluxoid}
\vec{E}(\vec{r}) = \lambda^2 {\rm rot}\vec{J_m}(\vec{r}) + \vec{\cal E}(\vec{r}).
\end{equation}
The integral form of this equation is nothing else but the fluxoid quantization.

One can obtain ${\vec E}(\vec{r})$ and ${\vec J_m}(\vec{r})$ in terms of the electric charge density $\rho(\vec{r})$ and the dual Dirac string ${\vec{\cal E}}({\vec r})$. The analytic solutions of eqs. (\ref{Gauss}), (\ref{Ampere}) and (\ref{fluxoid}) take the form

\parbox{11cm} {\begin{eqnarray*} \label{fields}
{\vec J_m}(\vec{r})&=&- {\rm rot}{\vec P}({\vec r}),\\
{\vec E}({\vec r})&=&{\vec P}({\vec r}) - {\rm grad}\phi ({\vec r}), \end{eqnarray*}} \hfill
\parbox{1cm}{\begin{eqnarray} \end{eqnarray}}

\noindent where

\begin{equation} \label{potential}
\phi ({\vec r}) =  \int \frac{d^3 r'}{4 \pi} \rho_e({\vec r}')
\frac{e^{-{\frac{1}{\lambda}}|{\vec r}-{\vec r}'|}}{|{\vec r}-{\vec r}'|}
\end{equation}
is the Yukawa potential of the sources and

\begin{equation} \label{polarisation}
{\vec P}({\vec r}) = \frac{e}{4 \pi \lambda^2} \int_{l} d \vec{r}' \quad \frac{e^{-\frac{1}{\lambda}|\vec{r}-\vec{r}'|}}{|\vec{r}-\vec{r}'|}
\end{equation}
describes the polarization of the vacuum. In ordinary electrodynamics polarization is given by the density of electric dipoles in the medium, in our case it is provided by monopole currents around the string. If we assume the string ${\vec{\cal E}}({\vec r})$ to be the straight line connecting the charges, the total energy of the electric field ${\vec E}({\vec r})$ can be calculated analytically as a function of the distance $d$. It contains a linearly rising and in addition a Yukawa potential:  
\begin{equation} \label{econt}
W (d) = \frac{1}{2} \int d^3r \vec{E}^2 ({\vec r}) = \frac{Q^2}{8 \pi \lambda^2} d -  \frac{Q^2}{4 \pi} \frac{e^{-\frac{d}{\lambda}}}{d} + {\rm a \ divergent \ constant}.
\end{equation}
It should be stressed that for a straight string of infinite length ($\vec{r}_+ \rightarrow -\infty, \vec{r}_{-} \rightarrow +\infty$) formulae (\ref{fields}) -- (\ref{polarisation}) restore the field and current distribution used in ref. \cite{haymaker}. The solution for the electric field takes the simple form ${E}(R) = (e / 2\pi\lambda^2) K_0(R/\lambda)$ in this case.

\begin{figure}
\centerline{\input{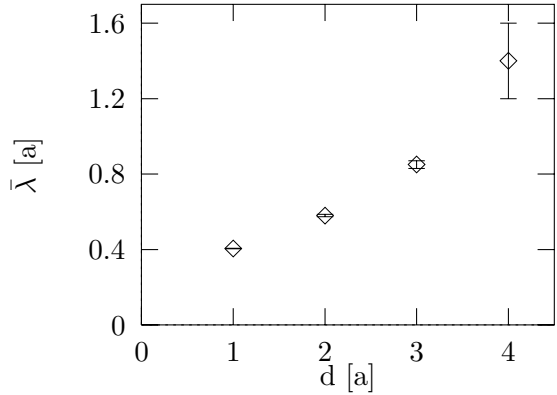}\hspace{1cm}}
\caption{\label{lamnd}The ``penetration length'' $\bar{\lambda}$, obtained by a fit of the electric field profile in the symmetry plane within the effective model considering just one straight string: The strong dependence on the distance of charges reflects the broadening of the flux tube.}
\end{figure}

\begin{figure}
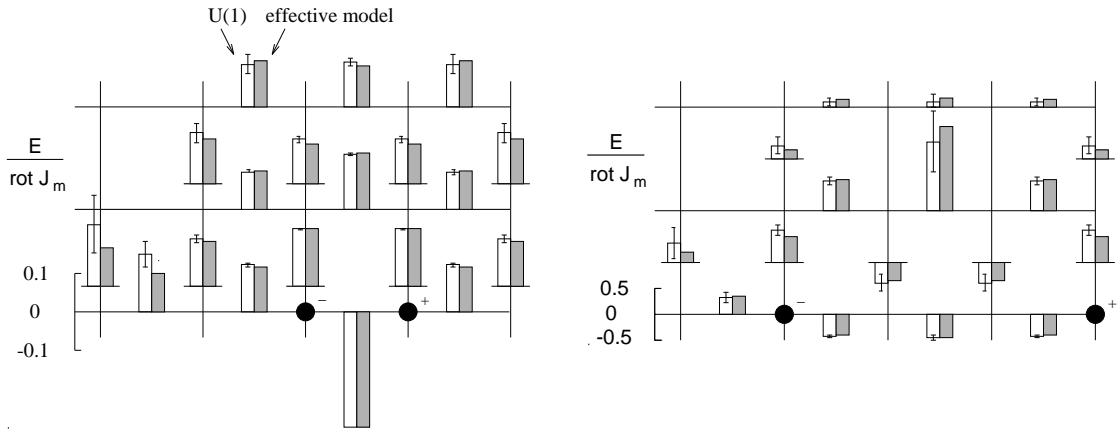

\centerline{\epsfxsize=6.95cm \epsfbox{ratio1.eps} \epsfxsize=7.85cm \epsfbox[-40 -58 370 182]{ratio3.eps}}
\vspace{1cm}
\caption{\label{ratio}On the links of the three-dimensional lattice we show the ratio of the electric field strength and the curl of monopole current in a plane of the $Q\bar{Q}$-axis for distances of charges $d=1a$ (left figure) and $d=3a$ (right figure): For each link the left box with errorbars shows the the result of $U(1)$ simulation, the right one shows the ratio obtained within the effective model with $\lambda=0.163 a$.}
\end{figure}

As discussed in the introduction we prefer to formulate the effective model of above equations on a lattice of the same size as we used for the Monte-Carlo calculations shown in Figs.~\ref{axis} to \ref{condensate}. On a spatial three-dimensional lattice the electric field resides on the links, whereas the monopole currents are located on plaquettes. By constructing the curl of these quantities one changes from links to plaquettes and vice versa. Therefore equ.~(\ref{fluxoid}) is defined on links, equ.~(\ref{Ampere}) on  plaquettes. The two coupled equations can easily be solved iteratively for arbitrary paths of the string ${\vec{\cal E}}$. For comparison to Monte-Carlo calculations of $U(1)$ lattice gauge theory we consider the fields and currents on the whole lattice and not only in the symmetry plane between the charges. Our results give clear evidence that in order to get reliable values for the penetration depth $\lambda$ it is necessary to consider fluctuations of the string $\vec{\cal E}$:
\begin{itemize}
\item If we try to fit $\lambda$ in the symmetry plane, using only a straight string, it turns out that an increasing distance between the charges from $d=1a$ to $d=4a$ would require strongly increasing $\lambda$-values $\bar{\lambda}(d)$, see Fig.~\ref{lamnd}, which in our case range from $0.4$ until above $1.0$. This is of course against physical expectations because the penetration depth $\lambda$ characterizing the confinement vacuum should be only a function of the coupling constant.
\item The $U(1)$ monopole current profiles for $d=3a$ and $d=4a$ show a minimum in the middle between charges (see Fig.~\ref{axis}). Within the effective model this cannot be explained if the string is just a straight line connecting the charges.
\item The ratio of the electric field strength and the curl of monopole current off the $Q\bar{Q}$-axis is far from being constant (see Fig.~\ref{ratio} for $d=3a$), as the validity of the London equation would suggest.
\end{itemize}

\begin{figure}
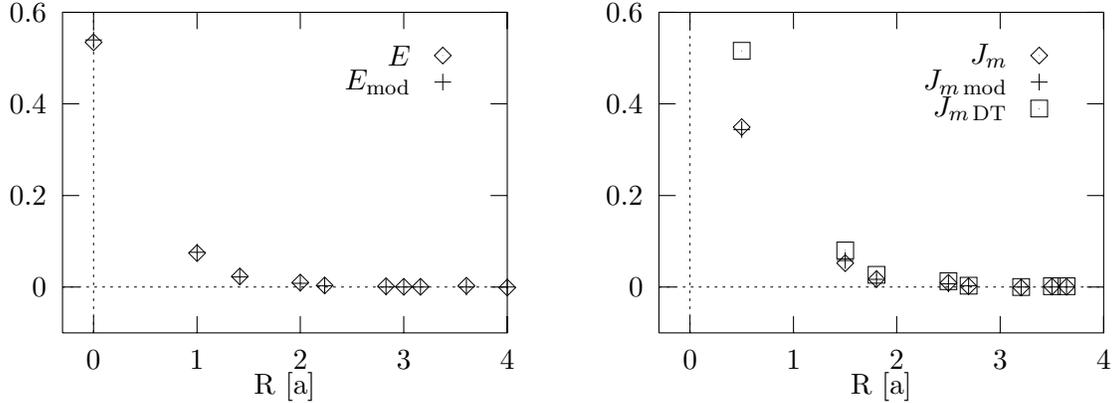

\centerline{\input{e_mod_nd2.tex}\input{jm_mod_nd2.tex}\hspace{1cm}}
\caption{\label{mod}Transverse flux tube profile ($d=2a$, zeta $=0.5a$) of electric fields and monopole currents in $U(1)$ ($E, J_m$) and in the effective model ($E_{\rm mod}, J_{m\,\rm mod}$) with the fitted parameter $\lambda=0.163a$. Errorbars have been omitted because they are smaller than the symbols. The monopole currents in DeGrand-Toussaint definition are also plotted ($J_{m\,\rm DT}$) for comparison. It turns out to be larger by a factor of approximately 1.5, independent on the strength of the signal.}
\end{figure}

Therefore, we perform a thermal average over some thousands of static string shapes, where strings up to a maximal length $l_{\rm max}=11a$ are considered. We use the same periodic boundary conditions as in $U(1)$ simulations, strings are therefore allowed to reach their final point over the period. The temperature of $U(1)$ simulations is $1/N_{t}a$, which means that in our simulation it is given in lattice units by $T=1/4$. Field configurations for a special path of the string have a classical energy 
\begin{equation} \label{eclass}
W = \frac{1}{2} \sum_{x, i=1,3} E_{x,i}^2,
\end{equation}
and contribute according to their Boltzmann weight $\exp(-W/T)$. A fit to all Monte-Carlo results of $\vec{E}$ and $\vec{J}_m$ on the whole lattice for the distances $d=1a$, $2a$, $3a$ and $4a$ gives $\lambda=0.163 a$ for $\beta=0.96$ with $\chi^2 \approx 1$. As an example for the excellent agreement we show a comparison of $U(1)$ and effective model data for the profiles of electric field and monopole current in Fig.~\ref{mod} (for a distance $d=2a$ between the charges). Because the statistical errors are very small for small distances a quantitative comparison is more significant in these cases. In the figure of the current profile we have also plotted the results of a measurement according to the DeGrand-Toussaint definition. It can be seen that the currents increase by a factor of around $1.5$ at the chosen $\beta$-value of $0.96$. Even for a single distance $d$ this definition would not allow to fit at the same time the exponential falling off and the absolute values of $\vec{E}$ and $\vec{J}_m$ by the effective model of coupled Maxwell-London equations (as long as $\sin \theta_{i4}$ is taken for measurement of electric field strength, which was shown to satisfy the Gauss law).

The ratio of the electric field strength and the curl of monopole current turns out to be very sensitive on string fluctuations. Within the effective model its strongly varying value can only be explained by considering fluctuations of the string where the London relation is replaced by the fluxoid quantization (as shown in Fig.~\ref{ratio}). Far from the charges this ratio should tend to $\lambda^2$ (which is difficult to check because of the decreasing signal in Monte Carlo calculations); in the vicinity of the charges it takes significantly larger values. This can be interpreted as increasing probability that the string runs along the regarded link. If the ratio gets negative the curl of the monopole current has changed its sign which indicates that the ``string'' contribution to the expectation value is dominating. For a distance $d=3a$ this happens not only on the line connecting the charges, but also for the transverse component near the symmetry plane (see Fig.~\ref{ratio}). The good agreement of this ratio between $U(1)$ and the effective model also shows that the classical energy (\ref{eclass}) is a good quantity for measuring the total energy for a single string shape. Remaining tiny deviations may be due to the fact that we have considered strings only up to a certain length. Strings closing periodically are therefore slightly underestimated. Another restriction of our effective model is the lack of time dependence of string fluctuations. 

\begin{figure}
\centerline{\input{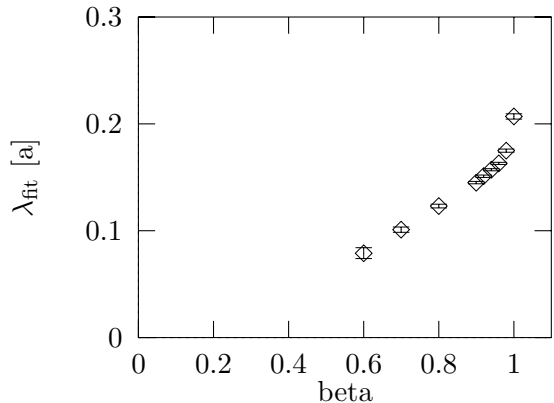}\hspace {1cm}}
\caption{\label{lam}The fitted penetration length $\lambda$ as a function of the inverse coupling $\beta$.}
\end{figure}

Finally we investigate the dependence of the fitted penetration length $\lambda$ on the inverse coupling $\beta$ (see Fig.~\ref{lam}). For this purpose we perform Monte-Carlo simulations for seven further values of $\beta$ from $\beta=0.6$ to $\beta=1.0$, where the phase transition to the Coulomb phase takes place. In the strong coupling regime the effective model works very well, because string fluctuations become less important. At $\beta \to 1.0$ the penetration length increases dramatically, which is in agreement with the physical expectation of a divergence of $\lambda$ at the phase transition from the ``superconducting'' to the normal state. Above the phase transition we are not able to describe the field and current distribution of $U(1)$ by means of our effective model which shows itself in strongly increasing $\chi^2$ values.   

\section{Conclusion}

In the confined phase of $U(1)$ lattice gauge theory electric field distributions and monopole currents around a pair of electric charges can be very well described  by Dirac's extension \cite{dirac} of Maxwell's electrodynamics supplemented by the dual London equation. The investigated properties of the highly non-linear $U(1)$ quantum system are equivalent to an effective model of coupled (linear) Maxwell and London equations with the penetration depth $\lambda$ as the only free parameter. A necessary condition for the agreement of both models for finite lattice constant $a$ is the identification of monopole currents in accordance with Maxwell equations. Further, it is essential to consider fluctuations of the dual Dirac string in the effective model. Proceeding this way the fields for all distances of charges can be described by a universal penetration depth $\lambda$. If the electric field penetrates into the monopole condensate monopole currents in time direction are suppressed and consequently the condensate is partly expelled from the region of electric fields. This leads to a repulsion between monopole condensate and electric flux lines and to the compression of the flux lines in a tube. Finally we would like to mention that the definition of monopole currents by the Maxwell equations can be generalized to non-Abelian gauge theories, allowing a gauge independent verification of the dual superconductor model in QCD.

\noindent With pleasure we acknowledge fruitful discussions with J. Greensite, A. Ivanov and \v S.~Olejn{\'\i}k.

\end{document}